\documentclass{aa}

\usepackage{textcomp}
\usepackage[varg]{txfonts}

\usepackage{graphicx}

\usepackage{natbib}
\bibpunct{(}{)}{;}{a}{}{,} 

\begin{document}

\title{Variations of dose rate observed by MSL/RAD in transit to Mars}

\author{Jingnan Guo\inst{\ref{inst:kiel}}
\and Cary Zeitlin\inst{\ref{inst:durham}}
\and Robert~F. Wimmer-Schweingruber\inst{\ref{inst:kiel}} 
\and Donald~M. Hassler\inst{\ref{inst:boulder}}
\and Arik Posner\inst{\ref{inst:nasa}}
\and Bernd Heber\inst{\ref{inst:kiel}}
\and Jan K\"ohler\inst{\ref{inst:kiel}}
\and Scot Rafkin\inst{\ref{inst:boulder}}
\and Bent Ehresmann\inst{\ref{inst:boulder}} 
\and Jan~K. Appel\inst{\ref{inst:kiel}} 
\and Eckart B\"ohm\inst{\ref{inst:kiel}} 
\and Stephan B\"ottcher\inst{\ref{inst:kiel}}
\and S\"onke Burmeister\inst{\ref{inst:kiel}} 
\and David E. Brinza\inst{\ref{inst:jpl}}
\and Henning Lohf\inst{\ref{inst:kiel}} 
\and Cesar Martin\inst{\ref{inst:kiel}} 
\and G\"unther Reitz\inst{\ref{inst:koeln}} }


\institute{Institute of Experimental and Applied Physics, Christian-Albrechts-University, Kiel, Germany\email{guo@physik.uni-kiel.de}\label{inst:kiel}
\and Southwest Research Institute, Earth, Oceans \& Space Department, Durham, NH, USA\label{inst:durham}
\and Southwest Research Institute, Space Science and Engineering Division, Boulder, USA\label{inst:boulder}
\and NASA Headquarters, Science Mission Directorate, Washington DC, USA\label{inst:nasa}
\and Jet Propulsion Laboratory, California Institute of Technology, Pasadena, CA, USA\label{inst:jpl}
\and Aerospace Medicine, Deutsches Zentrum f\"ur Luft- und Raumfahrt,   K\"oln, Germany\label{inst:koeln}}

\date{Received 14 January 2015/ Accepted date }

\abstract
{}
{To predict the cruise radiation environment related to future human missions to Mars, the correlation between solar modulation potential and the dose rate measured by the Radiation Assessment Detector (RAD) has been analyzed and empirical models have been employed to quantify this correlation.}
{The instrument RAD, onboard Mars Science Laboratory's (MSL) rover Curiosity, measures a broad spectrum of energetic particles along with the radiation dose rate during the 253-day cruise phase as well as on the surface of Mars.
With these first ever measurements inside a spacecraft from Earth to Mars, RAD observed  the impulsive enhancement of dose rate during solar particle events as well as a gradual evolution of the galactic cosmic ray (GCR) induced radiation dose rate due to the modulation of the primary GCR flux by the solar magnetic field, which correlates with long-term solar activities and heliospheric rotation. }
{We analyzed the dependence of the dose rate measured by RAD on solar modulation potentials and estimated the dose rate and dose equivalent under different solar modulation conditions. These estimations help us to have approximate predictions of the cruise radiation environment, such as the accumulated dose equivalent associated with future human missions to Mars.}
{The predicted dose equivalent rate during solar maximum conditions could be as low as one-fourth of the current RAD cruise measurement. However, future measurements during solar maximum and minimum periods are essential to validate our estimations.}


\maketitle
\titlerunning{Variation of MSL/RAD dose rate in transit to Mars}
\authorrunning{GUO ET AL.}


\section{Introduction to MSL/RAD}\label{sec:intro}
The Mars Science Laboratory (MSL) spacecraft, containing the Curiosity rover, was launched on November 26, 2011 and landed on Mars  on August 6, 2012 after a 253-day cruise. The Radiation Assessment Detector \citep[RAD; ][]{hassler2012}) onboard Curiosity provided the first ever radiation measurements during its cruise from Earth to Mars \citep{zeitlin2013} and now on the surface of the planet \citep{hassler2014} by measuring a broad spectrum of energetic particle radiation. The assessment of the radiation environment is fundamental for planning future human missions to Mars and evaluating the biologic effects of the radiation likely to be encountered on these kinds of missions.

\begin{figure}
\centerline{\includegraphics[width=0.6\columnwidth]{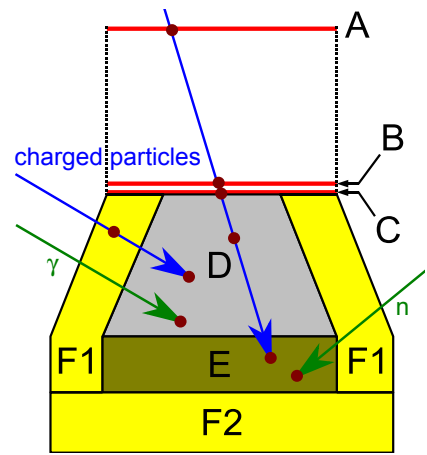}}
\caption{Schematic view (not to scale) of the RAD instrument, consisting of three silicon detectors (A, B, C, each having the thickness of 300 \textmu m), a caesium iodide scintillator (D with a height 
of 28 mm), and a plastic scintillator (E with a height of 18 mm). Both scintillators are surrounded by a plastic anticoincidence (F1 and F2). For detecting charged particles (blue), A, B, C, D, and E are used as a telescope. Neutral particles (both neutrons and gammas, green) are detected in D and E using C, F1, and F2 as anticoincidence. This figure is adapted from Figure 7 of \cite{koehler2011} and originally from Figure 3 of \cite{posner2005}.} 
\label{fig_RAD}
\end{figure}

RAD has detected galactic cosmic rays (GCRs) and solar energetic particles (SEPs), as well as secondary neutrons and other secondary particles created by the primary particles interacting with the spacecraft shielding. 
In a compact, low-mass, low-power instrument, RAD's innovative design combines charged- and neutral-particle detection capabilities over a wide dynamic range. A schematic of the RAD sensor head (RSH) is shown in Figure \ref{fig_RAD}. A detailed overview of the detectors is given by \citet{hassler2012}. 
For charged particle detection,  silicon detectors A and B are used in coincidence to define the acceptance angle of incoming particles. RAD measures mainly downward fluxes on the surface of Mars and it separates the charged particles in two categories: (a) particles stopping inside one of the detectors (B, C, D, or E), and (b) particles penetrating through the whole instrument and still creating energy depositions at the bottom detector of the sensor head. 
The total energy, charge, and mass of a stopping particle can be determined thus enabling its spectra to be reconstructed. 
The energy of a penetrating particle cannot, by definition, be easily recovered. However, the particle type and broad information about its energy can be retrieved using the particle's energy deposit pattern through the stack of detectors. The charged particle spectra measured with RAD on the surface of Gale crater are presented in \cite{ehresmann2014}. 
For neutral particle detection, both D and E detectors are, to different degrees, sensitive to $\gamma$-rays and neutrons. The separation and reconstruction of $\gamma$-ray and neutron energy spectra on the Martian surface measured by RAD have been accomplished using inversion methods \citep{koehler2014}.

\section{Radiation environment and dose rate variations in transit to Mars}\label{sec:radiation_cruise}
In interplanetary space, GCRs contain 98\% atomic nuclei and 2\% electrons, and the nuclei consist of about 87\% protons, 12\% helium, and only 1\% heavier nuclei \citep{simpson1983}. This composition however varies slightly depending on the phase of the heliosphere modulation, which is related to both the 11-year solar activity cycle and the heliospheric rotation\footnote{The Sun has a latitude-dependent sidereal rotation period between 25.4 days near the equator and 30+ days at the poles.}.
Broadly speaking, the stronger the solar activity the higher would be the deflection of interplanetary GCRs, resulting in a reduction of the GCR flux. This relationship can be often observed as an anticorrelation between neutron monitor counts (reflecting the GCR flux intensity at Earth) and the sunspot number (a standard measurement of solar activity) \citep{heber2007}.
This modulation is however energy-dependent: the flux of low-energy GCRs is more affected than that of the high-energy GCRs.     

On its journey to Mars, RAD started operating on December 6 2011 and, with few interruptions, measured through July 14 2012 when it was temporarily switched off in preparation for the landing.
RAD collected more than seven months of first-time measurements of the radiation environment inside the shielding of the spacecraft from Earth to Mars, in principle, a similar environment to future manned missions to the planet, thus providing insight into the potential radiation hazard for astronauts \citep{zeitlin2013}. 
Because of the shielding of the spacecraft and internal structures, RAD measured a mix of primary and secondary particles. The latter are produced by primary particles via nuclear or electromagnetic interactions as they traverse the spacecraft.  
A simplified shielding model of the spacecraft developed at JPL has been be used to calculate the shielding distribution as seen by RAD, which is mounted to the top deck of the rover \citep{zeitlin2013}.
Shielding around the RAD instrument during cruise was complex: most of the solid angle was lightly shielded with a column density smaller than 10 $\rm{g/cm^2}$, while the rest was broadly distributed over a range of depths up to about 100 $\rm{g/ cm^2}$. 

RAD measures dose in two detectors: the silicon detector B and the plastic scintillator E \citep{hassler2012}. The latter has a composition similar to that of human tissue and is also more sensitive to neutrons than silicon detectors. 
The bottom panel in Figure \ref{fig_cruise} shows the dose rate measurements taken by both detectors (black for E and gray for B) from 9 December 2011 to 14 July 2012 during the cruise to Mars. For a given incident flux, the dose rate in silicon is generally less than the dose rate in plastic because of the comparatively larger ionization potential of silicon.
There were generally 44 dose rate measurements per day and we have calculated the average values of the dose rate in each day, shown in the figure as red and magenta dots for detector E and B, respectively. 
During quiet periods in solar activity, the average GCR dose rate measured by the silicon and plastic scintillators was about 332 $\pm$ 23 \textmu Gy/day and 461 $\pm$ 92 \textmu Gy/day, respectively \citep{zeitlin2013}. 
In comparison with the measured dose rate in plastic, the dose rate in silicon can be converted using a conversion factor of 1.38 \citep{zeitlin2013},  which approximately relates energy loss per unit of path length ($\text{d}E/\text{d}x$) in silicon to linear energy transfer (LET) in water. 
After conversion, the dose rate in water, as measured in the silicon detector B, was 458 $\pm$ 32 \textmu Gy/day, which is comparable to measurements in the plastic scintillator within uncertainties \footnote{Subsequent to \citet{zeitlin2013}, we have re-evaluated the conversion factor used to obtain tissue dose from dose in silicon, and this yields the slightly lower dose rate (458 \textmu Gy/day) shown here compared to the earlier reported value (481 \textmu Gy/day). This is also the case for dose equivalent rate as shown later.}.

The dose rate during the few SEP events (marked as the yellow-shaded areas) were as high as \textgreater 10000 \textmu Gy/day. For a close visualization of the variations of the GCR-driven dose rate during the solar quiet periods, we omit the peak values of the SEPs in this figure. A zoomed-out figure containing the cruise dose rate measurements, as well as the peak doses of the SEPs can be found in \cite{zeitlin2013}. 

\subsection{Neutron monitor count rates and solar modulation strength}\label{sec:neutron_monitor}
Neutron monitor count rates probe the intensity of primary cosmic rays above certain thresholds in energy and magnetic rigidity, and therefore the measured flux depends on the location and sensitivity of the monitor.
A commonly used parameter of heliospheric modulation is the modulation potential $\Phi$ \citep{gleeson1968solar, usoskin2005}, which corresponds to the mean energy loss of a cosmic ray particle inside the heliosphere and is often used to parametrize the modulation of the GCR spectrum. The parameter 
$\Phi$ is defined in a spherically symmetric and steady-state case for the Earth's orbit.
\cite{usoskin2002}  used the following method to estimate $\Phi$ using the measured neutron monitor count rates: first, a stochastic simulation is used to calculate the galactic cosmic ray spectra at the Earth's orbit for different values of $\Phi$; second, these spectra are convoluted with the specific response function of the neutron monitor to generate the expected neutron monitor count rates given a certain value of $\Phi$; and finally, the modulation strength $\Phi$ can be derived inverting this forward-modeled relation from the actually recorded neutron monitor count rates. An estimated function of derivation of $\Phi$ (in the unit of MV) from the measured count rates ($CR_{NM}$, in the unit of counts per minute) is given as the following \citep{usoskin2002}: 
\begin{eqnarray}\label{eq:NMc2phi}
\Phi = A + \frac{B}{CR_{NM}^2},
\end{eqnarray}
where A and B are modeled values that depend on the geographical coordinates and hence the geomagnetic cutoff rigidity of the monitor stations.  

\begin{figure}
\centerline{\includegraphics[width=1.0\columnwidth]{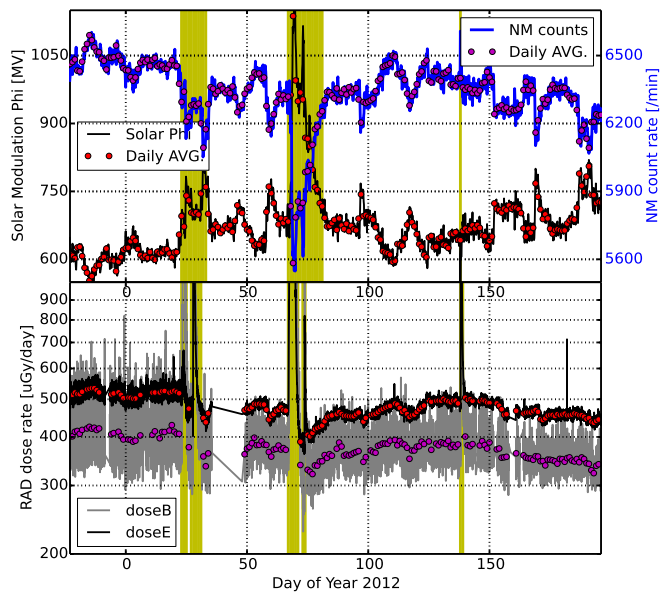}}
\caption{\textit{\textup{Top}}: blue line shows the Oulu neutron monitor count rate recorded during the time of MSL's cruise to Mars and the daily averaged values are shown as magenta dots. The black line shows the solar modulation potential $\Phi$ derived from the neutron monitor count rates and its daily average is shown as red dots. The RAD observed SEPs are highlighted in yellow shaded areas in which the data are not considered in the correlation study. 
\textit{\textup{Bottom}}: dose rate recorded by RAD in silicon detector B (gray curve) and in plastic scintillator E (tissue-equivalent, black curve). Their daily averaged values are plotted as solid dots in magenta and red. 
}\label{fig_cruise}
\end{figure}

\citet{posner2013} have shown that during most of the cruise phase the MSL spacecraft was well connected with the Earth via the Parker spiral magnetic field, a circumstance now referred to as the Hohmann-Parker effect. As a consequence, it was shown that the cosmic ray background at MSL and the Solar and Heliospheric Observatory (SOHO) near Earth were highly correlated. Therefore, for the evaluation of the solar modulation of GCR-induced MSL/RAD measurements, we can  take  approximately the modulation potential calibrated from measurements of the Oulu Neutron Monitor on Earth. 
The top panel of Figure \ref{fig_cruise} shows both the count rates (blue, in the unit of counts per minute) recorded by the Oulu neutron monitor \footnote{The Oulu count rate data were obtained from http://cosmicrays.oulu.fi/ and the pressure effect was corrected.} and the calibrated solar modulation potential $\Phi$ (black, in the unit of MV). We opted to use Oulu data with a resolution of three hours and for the current variation study, we calculated the daily average values of both neutron count rates and $\Phi$ shown as solid dots in magenta and red, respectively. The daily variations of the measurements are folded into the sigma errors of the daily-average values.

\subsection{Correlation between  solar modulation and  cruise dose rate}\label{sec:correlation}
Impacts of solar events are often observed in the form of Forbush decreases in the GCR flux due to the enhancement of the interplanetary magnetic field sweeping away more GCRs \citep{forbush1938, zeitlin2013}. During SEPs and Forbush decreases, when the GCR spectrum is distorted by the nonpotential enhanced solar fields, the evaluations of $\Phi$ bear large uncertainties.
Several Forbush decreases and SEPs detected by the Oulu neutron monitor and MSL/RAD are highlighted as yellow-shaded areas in Figure \ref{fig_cruise}. 
They do not necessarily match precisely in time because of the spatial variability and time delay of solar events passing the two instruments. 
For the current correlation study of GCR-induced dose rate and solar modulations, we exclude the data collected during these five highlighted solar events since these measurements are due to both modulated GCRs and SEPs. 
During the quiet period, 
we calculated the daily averaged neutron count rate $CR_{NM}$, modulation potential $\Phi$, as well as the daily averaged RAD dose rate shown in the bottom panel
of Figure \ref{fig_cruise} as red and magenta dots for E and B detector, respectively. 
The sigma error of each daily average value is taken to be the standard deviation of all the measurements within that day. 
The daily average data clearly indicate an anticorrelation between the variation of the RAD dose rate and the solar modulation potential $\Phi$. The correlation coefficient between two variables can be obtained as the covariance of the two variables divided by the product of their standard deviations, giving a value between +1 and -1 inclusive, where 1 is total positive correlation, 0 is no correlation, and -1 is total negative correlation \citep[e.g.,][]{stigler1989}.  
We obtained the correlation coefficient between the daily averaged solar modulation $\Phi$ and the RAD dose rate (during periods with no SEPs) to be -0.80 for detector B and -0.77 for detector E, indicating a very good negative correlation between the solar modulation and the measured GCR induced dose rate. 

The radial gradient of galactic cosmic rays in the inner heliosphere has been studied by \citet{gieseler2008radial} and they estimated GCR flux increases at a radial gradient rate of 4.5\% per AU in the inner heliosphere. Considering that the spacecraft moved progressively outward in the solar system from 1 AU to about 1.5 AU, we  corrected this gradient increase on the measured dose rate. This effect is however very small (within 2\% of correction) and   has only  affected slightly the analysis that follows.

A quantitative study of the anticorrelation is challenging because of the lack of analytic models. 
We employed two separate empirical models to describe this anticorrelation. 
\begin{enumerate}
\item A simple linear correlation described by a first order polynomial function is used to fit the daily average data, i.e., 
\begin{eqnarray}\label{eq:Correlation1}
D = \beta_0 + \beta_1 \Phi,  
\end{eqnarray}
where D is the RAD measured dose rate (either by detector B or E) in the unit of \textmu Gy/day and $\beta_1$ in the unit of \textmu Gy/day/MV represents the linear relation between the changing of the solar modulation and the variance of observed dose rate. 
This linear correlation can be easily fitted by a simple regression method, which  fits a straight line through the set of measurements minimizing the sum of squared residuals (i.e., the chi square function) of the model. Sigma errors of the data can also be included in the fitting as a weighting factor. The fitted parameters and their standard deviations are $\beta_0 = 626 \pm 48$ \textmu Gy/day and $\beta_1 = -0.39 \pm 0.07$ \textmu Gy/day/MV for dose rate in B and $\beta_0 = 758 \pm 27$ \textmu Gy/day and $\beta_1 = -0.44 \pm 0.04$ \textmu Gy/day/MV for dose rate in E. 
Given the small error bars of the fitted parameters ($< 15\%$), a linear function seems to be a good approximation for the range of the measured modulation, i.e., $\Phi$ between 550 MV and 800 MV. 
 
\item We employ an empirical model, which is often used in the analysis of neutron count rates \citep[e.g.,][]{usoskin2011}, i.e., 
\begin{eqnarray}\label{eq:Correlation-1}
D = \alpha_0 + \frac{\alpha_1} {\Phi + \alpha_2}, 
\end{eqnarray}
with the requirement of $\alpha_2 \geq 0 $. Otherwise, $\Phi + \alpha_2 = 0$ would lead to nonphysical solution of the problem. Also by minimizing the chi square function, we fitted the above function and obtained the parameters as 
$\alpha_0 = 94.4 \pm 2.6$ \textmu Gy/day, $\alpha_1 = (1.8 \pm 0.02) \times 10^5 $ MV $\cdot$ \textmu Gy/day and $\alpha_2 = 0.0002 \pm 1.0 $ MV for dose rate in B and $\alpha_0 = 180.6 \pm 0.7$ \textmu Gy/day, $\alpha_1 = (1.9 \pm 0.005) \times 10^5 $MV $\cdot$ \textmu Gy/day and $\alpha_2 = 0.0008 \pm 1.4 $ MV for dose rate in E. 
The error bars for both $\alpha_0$ and $\alpha_1$ are below 3\% while the relative error for $\alpha_2$ is much bigger because of the small absolute values of $\alpha_2$, which indicates that the parameter $\alpha_2$ could be probably ignored in the model fitted by the current measurement.
\end{enumerate}

\subsection{Estimates of the cruise radiation environment under different solar modulation conditions}\label{sec:extrapolation}
\begin{figure}
\centerline{\includegraphics[width=1.0\columnwidth]{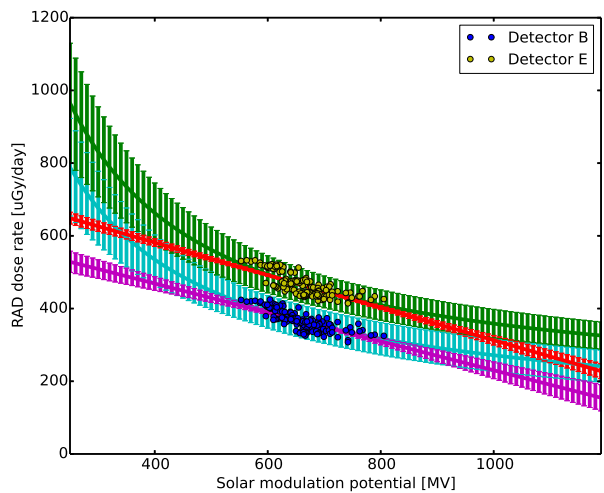}}
\caption{Data and fittings processed through bootstrap Monte Carlo simulations of the variations of the RAD dose rate and the solar modulation $\Phi$ during the cruise to Mars.
The red/magenta lines represent the linear fits (Eq.~\ref{eq:Correlation1}) of RAD measured dose rate in detector E/B versus solar modulation potential. The green/cyan lines show the results of the nonlinear fits (Eq.~\ref{eq:Correlation-1}).}\label{fig_cruise_fit}
\end{figure}

It is difficult to derive analytically whether the above models would serve as good approximations when solar modulation $\Phi$ is outside the range between 550 and 800 MV of the studied case. 
In general, solar modulation affects particles with different energies in different degrees, i.e., $\Phi$ has a stronger/weaker effect on GCR particles with lower/higher energies \citep{heber2007}. For any given species, a lower-energy ion contributes more to dose than does a higher-energy ion; we therefore expect that solar modulation has a more pronounced effect on dose than it does on integral fluxes.
Furthermore, the RAD dose rate during cruise was measured under the spacecraft shielding, which is more effective against lower energy particles that are more influenced by the variation of $\Phi$. 
A more detailed quantitative analysis could be possibly carried out in the future using a GEANT4 Monte Carlo simulation \citep{agostinelli2003} considering the energy-dependent solar modulations and the energy-dependent detector responses as well as the complex shielding model of the spacecraft. 
Nevertheless, based on our empirical regression fittings of the above two models, we could extrapolate our results to a wider range and make rough estimations of the dose rate values under different solar modulation conditions. 

A simple error propagation method based on the fitted parameters and their errors is however not reliable when extrapolating the function away from the measured data. To derive the robust estimates of the uncertainties due to both measurements and extrapolations, we have carried out a bootstrap Monte Carlo simulation \citep{efron1981nonparametric}. 
Simulated data sets are generated using the uncertainty range of the measured data, which are then used for fitting the model (according to either Eq.~\ref{eq:Correlation1} or Eq.~\ref{eq:Correlation-1}) and extrapolating to a wider range of $\Phi$  to obtain an accurate estimate of the uncertainties in the extrapolated data. 

The regression fittings of the above two models and the extrapolations of the fittings (over a range of $\Phi$ from 250 to 1200 MV), using bootstrap Monte Carlo simulations, are shown in Figure~\ref{fig_cruise_fit}.
As expected, the uncertainty of the extrapolated dose rate increases when the extrapolation is further away from the actual measurements.  
The differences of dose rate between two models are bigger for extrapolated points, especially for small solar modulation, i.e., solar minimum case. For instance, at $\Phi = 250 MV$ , the discrepancy between the two models is as large as 300 \textmu Gy/day for the plastic detector E.   
Nonetheless, preference of a better model is difficult to reach because of  the  limited statistics and range of the current data set. 
Future measurements over a wider range of solar modulations are necessary prior to human flights to Mars  to help improve and determine the models.

\subsection{Predictions of the dose equivalent rate}\label{sec:dose_equivalent}
\begin{figure}
\centerline{\includegraphics[width=1.0\columnwidth]{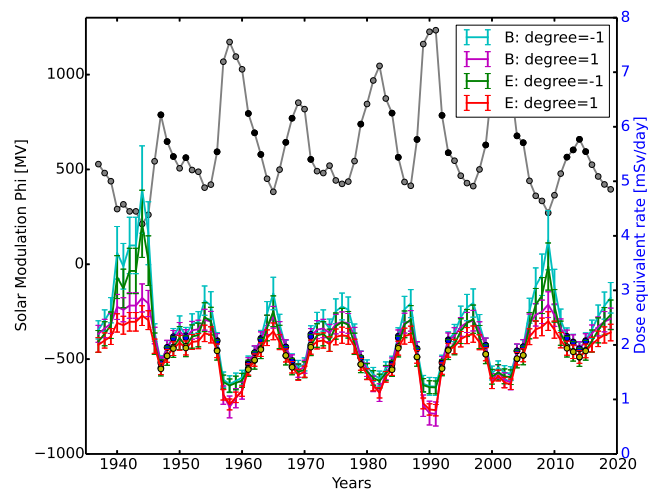}}
\caption{Annual average of reconstructed modulation potential $\Phi$ since 1937 (gray line and dots with units on the left axis) and estimated annual dose equivalent rate derived from different models (with units on the right axis). Cyan/Green line shows the dose equivalent values derived from the nonlinear model (Eq.~\ref{eq:Correlation-1}) based on detector B/E measurements. Magenta/Red line represents the dose equivalent values derived from the linear model (Eq.~\ref{eq:Correlation1}) based on detector B/E measurements. Black, yellow, and blue dots represent the data within the range of modulation $\Phi$ between 550 and 810 MV which the RAD cruise measurements covered.}\label{fig_cruise_extrapolated_dose}
\end{figure}

The Oulu neutron monitor \citep{usoskin2011}  published the annual values of reconstructed modulation potential $\Phi$ since 1937 and, accordingly, we are able to calculate the extrapolated dose rate values.
We  also correlated the monthly $\Phi$ values and  monthly sunspot number provided by the Royal Observatory of Belgium, Brussels (http://sidc.oma.be/silso/datafiles) using the same method as described in the last section.
Based on the sunspot predictions in the next five years (2015 to 2019) prepared by the U.S. Dept. of Commerce, NOAA, Space Weather Prediction Center (SWPC), we estimate the corresponding modulation potential as shown as the last five points in Figure~\ref{fig_cruise_extrapolated_dose}. 

Dose equivalent, in Sieverts (Sv), defined by the International Commission on Radiological Protection, is taken to be proportional to the risk of lifetime cancer induction via populations studies \citep[e.g.,][]{zeitlin2013}. It can be approximately estimated to be <$\rm{Q}$> $\times$ D, where <$\rm{Q}$> is the average quality factor, which is a conventional parameter for radiation risk estimation. A value of 3.82 $\pm$ 0.25 was measured during MSL's cruise phase \citep{zeitlin2013}.
Multiplying the measured tissue dose rate with the average quality factor yields the GCR dose equivalent rate. For the dose rate measured by detector B, a silicon-water conversion factor of 1.38 has to be applied first as described in Section~\ref{sec:radiation_cruise}.  

The extrapolated dose rate estimations at different modulation potential values can therefore be used to evaluate the dose equivalent rate as shown in Figure~\ref{fig_cruise_extrapolated_dose}. 
The annual average of reconstructed modulation potential $\Phi$ shows a clear 11-year cycle following the solar cycle and it varies over a wide range from less than 250 MV to more than 1200 MV; 
the estimated annual dose equivalent rate ranges from about 0.5 mSv/day to about 5 mSv/day and has a clear anticorrelation with $\Phi, $ as expected from both of our models (Eqs.~\ref{eq:Correlation1} and \ref{eq:Correlation-1}); 
stronger solar modulations lead to a decrease of dose equivalent rate and, at $\Phi >$ 1000 MV, the dose equivalent rate can be as low as $<$ 1 mSv /day within the uncertainties, approximately half of the average RAD cruise measurement \citep{zeitlin2013}: 1.75 $\pm$ 0.30 mSv/day; 
at solar maximum and minimum when the modulation potential is further away from the measured range, the discrepancy between the two models increases: the linear model often predicts a smaller dose rate than the nonlinear model as also shown in Figure~\ref{fig_cruise_fit}; 
large uncertainties emerge at low solar modulation potentials especially for the nonlinear model and future measurements over solar minimum periods prior to human flights to Mars are essential for improving the predictions at low modulation potentials.

\section{Discussions and conclusions}
We presented the dose rate data collected by MSL/RAD during its cruise phase from Earth to Mars and analyzed its variations during solar quiet periods. The variation of the GCR-induced dose rate at MSL occurs nearly simultaneously with the variation in neutron monitor count rate at Earth and was mainly driven by changes in heliospheric conditions. The simultaneity of cosmic ray count rate changes at MSL and Earth is attributed to Hohmann-Parker effect conditions as described in \citet{posner2013}. We show here that dose rate at MSL during the Mars transfer can be described by a solar modulation factor ($\Phi$) calculated from the neutron monitor count rate on Earth.
The results show a clear anticorrelation trend, with a correlation coefficient of $\approx$ -0.80 between $\Phi$ and the measured dose rate, indicating that when the solar activity is stronger GCR particles are attenuated more resulting in a decrease in the dose rate measurements.
The measurement from DOSIS onboard ISS has also shown an anticorrelation between solar modulation and the dose rate in the silicon detector (through personal communications with S\"onke Burmeister). However, the results are not directly comparable with RAD's measurement because ISS is in low Earth orbit (LEO) where Earth itself and its magnetic field provide shielding against GCRs. 

A quantitative study of this anticorrelation has been carried out employing two empirical models, which fit the data equally well as shown by the small standard deviation of the fitted line.
We further extrapolated the regression fittings to a wider range of solar modulations  to make approximate empirical predictions of dose rate and dose equivalent rate under different modulation conditions.
We employed a bootstrap Monte Carlo simulation  to include possible propagated errors due to the uncertainties of both measurements and extrapolations. The final derived dose equivalent rate at various modulation potential values are shown with a clear anticorrelation. The predicted dose equivalent rate during solar maximum conditions, where $\Phi \sim 1200$ MV was found to be as low as $\approx$ 0.5-1.5 mSv/day (as within the uncertainties of both models), which is considerably lower than the RAD cruise measurement $1.75 \pm 0.30$ mSv/day. The discrepancy of the two fitted models, as well as the extrapolation uncertainties, are significant at low solar modulation potentials and future measurements during solar minimum periods are necessary for improving the predictions at this range.
 
We also estimated the solar modulation potential in the next five years, using its correlation with sunspot numbers and employing the same method of correlating and extrapolating dose rate and $\Phi$. As the predicted solar modulation will decline in the future years, we predict a trend of increasing dose equivalent rate (between 1.7 and 3 mSv/day) from year 2015 until 2020. 

Total mission GCR dose equivalent can be estimated based on our measurements and predictions.
Considering a similar shielding condition, assuming a 180-day, one-way duration as a typical NASA’s “Design Reference” Mars mission \citep{drake2010}, we could estimate a crew taking a 360-day round trip to receive about $360 \pm 180$ mSv (the error bar combines the sigma from both models) from GCRs under a high solar modulation condition ($\Phi \approx$ 1200 MV).
The fastest round trip with on-orbit staging and existing propulsion technologies has been estimated to be a 195-day trip (120 days out, 75 days back with an extra, e.g., 14 days on the surface), as described by \citet{folta2012fast}. This would result in an even smaller GCR-induced cruise dose equivalent during solar maximum: $195 \pm 98$ mSv.

However, this fast round-trip option requires extra propulsion, which may result in a reduction of payload mass and a change of  shielding conditions. It is difficult to estimate how  reduced shielding would affect the dose rate, but it is probable that particles would traverse through more easily, heavy ions less likely to break up into lighter ions, and the production of secondary particles, such as neutrons, reduced. The dose rate and the quality factor are expected to be modified, both folding into the changes of dose equivalent. 
Monte Carlo GEANT4 simulations could be carried out for investigating this complicated process where realistic shielding geometry and materials can be included.   
In the case of the MSL spacecraft, the shielding condition around RAD was highly nonuniform \citep{zeitlin2013}: nearly 30\% of incoming particle trajectories within the field of view of RAD from above (coincidence of detector A and B in Figure~\ref{fig_RAD}) were only lightly shielded ($<$ 5 g cm$^{-2}$), the 50\% fraction from below were shielded by 8-10 g cm$^{-2}$, and the remaining 20\% were heavily shielded ($\geq$ 20 g cm$^{-2}$), giving an average depth of roughly 16 g cm$^{-2}$. 
It is highly probable that the crewed mission would not benefit from a perfectly uniform shielding. Also, the shielding materials for MSL have been rather complex: the spacecraft has a mix of materials that included hydrazine, mylar (parachute material), as well as aluminum and carbon composite (spacecraft shell).

However, we obtained the correlation between solar modulation potential $\Phi$ and dose rate    based on a very good magnetic connection between Earth and the spacecraft, allowing the direct application of the Neutron Monitor data collected at Earth. In the case of a weak magnetic connection between Earth and the spacecraft such as in a rapid Mars transit scenario, the local solar modulation potential $\Phi$ at the spacecraft location should be considered. Our current analysis  simply assumes that the solar modulation measured at Earth is representative for that at the Earth orbit and we  include the radial gradient of this modulation. Future work may also consider the longitudinal difference of the modulation across the Parker spirals.

It should also to be stressed that additional contributions of dose rate and dose equivalent rate by SEPs can differ significantly from our current measurements since the frequency and intensity of SEP events are highly variable. 
The MSL/RAD data were obtained under the rising and maximum phase of an unusually weak solar cycle. Only five SEP events were observed during the 253-day cruise to Mars as already reported by \citet{zeitlin2013}: two from 23 to 29 January with total dose equivalent being 4.0 mSv, two from 7 to 15 March with total dose equivalent being 19.5 mSv, and one on 17 May of 2012 with total dose equivalent being 1.2 mSv. The summed dose equivalent of the five observed SEPs is 24.7 mSv, roughly 5\% of the contribution to the total cruise dose equivalent ($\approx$ 466 mSv). 
During the peak of a solar maximum, where the solar activity is much more intensive, one would generally expect a bigger contribution of SEP-related dose equivalent. Nonetheless, a shorter trip (360-day or 195-day return) would also reduce the risk of encountering many SEP events. If we assume that on a 180-day trip during solar maximum peak a crew would receive four times the total SEP-contributed dose equivalent, i.e., $\approx$ 100 mSv, a round trip to Mars would result in a total dose equivalent of $560 \pm 180$ mSv, still smaller than the estimation based on the current solar modulation conditions, i.e., 662 $\pm$ 108 mSv \citep{zeitlin2013, hassler2014}. A 195-day round trip would of course further reduce our dose level estimations. 
These estimations are slightly less than the safe upper limit for 30- to 60-years old, nonsmoking females (600-1000 mSv) and males (800-1200 mSv) given by the Central estimates of dose limits \citep{cucinotta2011updates}.
However, given the uncertainties in the models, extra caution would be necessary in planning a human mission during solar maximum. Thus, future measurements over maximum solar modulation potentials are essential to refine our predictions of the reduced dose rate at this range.

%

%
%
%
%
%

\begin{acknowledgements}
RAD is supported by the National Aeronautics and Space Administration (NASA, HEOMD) under Jet Propulsion Laboratory (JPL) subcontract \#1273039 to Southwest Research Institute and in Germany by DLR and DLR's Space Administration grant numbers 50QM0501 and 50QM1201 to the Christian Albrechts University, Kiel. Part of this research was carried out at JPL, California Institute of Technology, under a contract with NASA. 
We thank Shawn Kang at JPL for his work on the shielding model of the spacecraft during the cruise phase.
The sunspot data has been obtained from Source: WDC-SILSO, Royal Observatory of Belgium, Brussels. We are grateful to the Cosmic Ray Station of the University of Oulu and Sodankyla Geophysical Observatory for sharing their Neutron Monitor count rate data. 
The data used in this paper are archived in the NASA Planetary Data System’s Planetary Plasma Interactions Node at the University of California, Los Angeles. The archival volume includes the full binary raw data files, detailed descriptions of the structures therein, and higher-level data products in human-readable form. The PPI node is hosted at the following URL: http://ppi.pds.nasa.gov/.
\end{acknowledgements}

%
\bibliographystyle{aa} 
\bibliography{msl_rad_guo} 

\end{document}